\documentclass[journal]{IEEEtran}

\usepackage{url}

\usepackage{caption}
\usepackage{hyperref}
\usepackage{amsmath}
\usepackage{amssymb}
\usepackage{mathtools}
\usepackage{adjustbox}
\usepackage{lipsum}
\usepackage{algorithm}
\usepackage{textcomp}
\usepackage{mathtools}
\usepackage{adjustbox}
\usepackage{lipsum}
\usepackage{stackengine}
\usepackage{graphicx}
\graphicspath{ {./images/} }
\usepackage{subcaption}

\usepackage{algpseudocode}
\usepackage{expl3}
\usepackage{float}
\usepackage{amsthm}
\usepackage{subfloat}
\newtheorem{thm}{Theorem}

\usepackage{tikz}
\usepackage{pifont}%
\usepackage{mwe}
\usepackage{cite}

\newcommand{\ie}{\textit{i}.\textit{e}., }
\newcommand{\eg}{\textit{e}.\textit{g}. }

\DeclareMathOperator{\diag}{\operatorname{diag}}

\DeclareMathOperator{\tr}{\operatorname{tr}}
\DeclareMathOperator{\Bernoulli}{\operatorname{Bernoulli}}

\begin{document}

\title{Kernel-based Joint Multiple Graph Learning and Clustering of Graph Signals}

\author{
Mohamad H. Alizade,
Aref Einizade,
and Jhony H. Giraldo
\thanks{M. H. Alizade (corresponding author) is with the Electrical Engineering Department of Sharif Uni. of Tech., Iran. E-mail: mhmd.h.alizade98@gmail.com.}
\thanks{A. Einizade and J. H. Giraldo are with LTCI, Télécom Paris, Institut Polytechnique de Paris, Palaiseau, France. E-mail: \{aref.einizade, jhony.giraldo\}@telecom-paris.fr.}
\thanks{This work was partially supported by the center Hi! PARIS.}}

\markboth{IEEE Signal Processing Letters, Vol. xx, No. y, MONTH YEAR}
{Shell \MakeLowercase{\textit{et al.}}: Bare Demo of IEEEtran.cls for IEEE Journals}
\maketitle

\begin{abstract}

Within the context of Graph Signal Processing (GSP), Graph Learning (GL) is concerned with the inference of the graph's underlying structure from nodal observations.
However, real-world data often contains diverse information, necessitating the simultaneous clustering and learning of multiple graphs.
In practical applications, valuable node-specific covariates, represented as kernels, have been underutilized by existing graph signal clustering methods.
In this letter, we propose a new framework, named Kernel-based joint Multiple GL and clustering of graph signals (KMGL), that leverages a multi-convex optimization approach. 
This allows us to integrate node-side information, construct low-pass filters, and efficiently solve the optimization problem.
The experiments demonstrate that KMGL significantly enhances the robustness of GL and clustering, particularly in scenarios with high noise levels and a substantial number of clusters. 
These findings underscore the potential of KMGL for improving the performance of GSP methods in diverse, real-world applications.

%The experiments demonstrate the algorithm's superiority over state-of-the-art approaches, highlighting the potential of node-side information to enhance the clustering and learning of graph signals in real-world scenarios.

% Within the context of Graph Signal Processing (GSP), Graph Learning (GL) is concerned with the inference of a graph's topology from nodal observations, \ie graph signals. 
% However, data is often in mixed form, relating to different underlying structures. This heterogeneity necessitates the joint clustering and learning of multiple graphs. In many real-life applications, there are available node-side covariates (\ie kernels) that imperatively should be incorporated, which has not been addressed by the rare graph signal clustering approaches. To this end and inspired by the rich K-means framework, we propose a novel kernel-based algorithm to incorporate this node-side information as we jointly partition the signals and learn a graph for each cluster. 
% Numerical experiments demonstrate its effectiveness over the state-of-the-art.

\end{abstract}


\IEEEpeerreviewmaketitle

\section{Introduction}

\IEEEPARstart{T}{he} emerging field of Graph Signal Processing (GSP) has introduced a plethora of analytical techniques \cite{shuman2013emerging,ortega2018graph,leus2023graph}. 
GSP focuses on the manipulation and analysis of data represented as signals associated with the nodes of a meaningful graph.
While some datasets, like traffic data, naturally exhibit graph-like structures, many others lack a known graph topology \cite{dong2016learning}. 
This has stimulated the growing popularity of Graph Learning (GL) within GSP \cite{dong2019learning,mateos2019connecting}.
GL encompasses various approaches, including those that employ physical processes like diffusion for data interpretation \cite{shafipour2021identifying,einizade2021simultaneous} and methods that assume neighboring nodes exhibit similar values, promoting global smoothness within the graph \cite{dong2016learning,kalofolias2016learn,guo2023graph,fatima2022learning}.

Previous GL methods have primarily dealt with homogeneous datasets, where the data is associated with a single graph \cite{dong2016learning,kalofolias2016learn}. 
However, many real-world datasets are heterogeneous, comprising clusters with diverse underlying structures. 
This heterogeneity results in the partition of graph signals, where each partition corresponds to a distinct, often unknown, graph.
For example, in fMRI datasets, brain imaging reveals various cognitive processes across different parts of the brain \cite{maretic2020graph}.
Each graph signal in such datasets may correspond to a separate cognitive process and, consequently, a distinct functional network \cite{saboksayr2021accelerated}.
Additionally, there is often node-specific information, such as spatial coordinates in a sensor network, or non-numeric data like categories or text \cite{pu2021kernel}.
Recent efforts have emerged to address the simultaneous GL and clustering of graph signals in these complex scenarios \cite{maretic2020graph,araghi2019k,karaaslanli2022simultaneous,yuan2022gracge}. 
Yet, none of these methods have effectively exploited node-side information to enhance their performance.
Moreover, they typically do not reconstruct the filtered (noiseless) graph signals, which represents a significant practical limitation.

In this letter, we introduce a new framework that combines node-specific information to simultaneously cluster the graph signals and learn the graph's underlying topology. 
To achieve this, we map the node-specific information into elements of a kernel's Hilbert space. 
The kernel matrix represents the covariates of the relationship between nodes in the Hilbert space, and we create low-pass filters by combining the Laplacian matrix with the inverse of the kernel matrices.
Thus, we introduce an iterative approach called the Kernel-based joint Multiple Graph Learning and clustering of graph signals (KMGL\footnote{\href{https://github.com/mohamad-h-alizade/KMGL}{https://github.com/mohamad-h-alizade/KMGL}}) algorithm, inspired by the K-means clustering framework \cite{ikotun2022k} and kernel-based GL methods \cite{pu2021kernel}.
To efficiently solve our optimization problem, we employ the Block Coordinate Descent (BCD) method \cite{beck2013convergence}. 
We link our GL task to a widely studied least squares problem, enabling us to optimize and solve our framework effectively.

Our work brings several contributions to the field:
i) This is the first study to incorporate node-specific information into multiple GL and signal clustering while also obtaining denoised graph signals.
ii) We demonstrate the convergence of the KMGL algorithm by exploiting the multi-convexity of the optimization problem.
iii) Our experiments reveal that leveraging node-specific information significantly enhances the robustness of GL and clustering, particularly when dealing with high levels of noise and a large number of clusters.
In contrast, existing methods struggle with severe performance deterioration.
iv) We extend our framework to handle cases where data is missing, providing further flexibility and practicality (see Appendix A for details).

\section{Preliminaries}

%\noindent \textit{Notation}: 

\subsubsection{Notation}

Vectors, matrices, and sets are denoted by boldface lowercase, boldface capital, and calligraphic capital letters, respectively.
The notations \((\cdot)^\top\), \(\tr(\cdot)\), \( \Vert \cdot \Vert _p\), and \(\Vert \cdot \Vert_F\) stand for the transpose operator, the trace operator, the \(p\)-norm of a vector, and the Frobenious norm of a matrix, respectively. The matrix \(\diag(\mathbf{a})\) is a diagonal matrix with the elements of the vector \(\mathbf{a}\) on its principal diagonal. 
The \((i,j)\)th and \(i\)th elements of a matrix \(\mathbf{M}\) and a vector \(\mathbf{x}\) are denoted as \(\mathbf{M}_{ij}\) and \(x_i\), respectively. The cardinality of set \(\mathcal{I}\) is stated by \(\vert\mathcal{I}\vert\).

\subsubsection{Graph Signals}

Let $\mathcal{G}=(\mathcal{V}, \mathcal{E}, \mathbf{W})$ be a weighted undirected graph without self-loops, where $\mathcal{V} = \{ v_1, \dots, v_n \}$ is the node set with $\vert \mathcal{V} \vert=n$, the edge set $\mathcal{E} \subset \mathcal{V} \times \mathcal{V}$, and $\mathbf{W}$ is the symmetric adjacency matrix.
The entity $\mathbf{W}_{ij}$ has a positive value if there is an edge between vertices $v_i$ and $v_j$ but zero otherwise. 
Let $\mathbf{D} = \diag(\mathbf{W}\mathbf{1})$ be the diagonal degree matrix, where $\mathbf{1}$ is the all-one vector of size \(n\).
%and $ D_{ii} $ denotes the summation of edge weights connected to node $v_i$. 
The Laplacian matrix of $\mathcal{G}$ given by $\mathbf{L} = \mathbf{D}-\mathbf{W}$ is symmetric and positive semi-definite with eigendecomposition $\mathbf{L}=\mathbf{U}\mathbf{\Lambda}\mathbf{U}^\top$ \cite{ortega2018graph}. 
%Here, $\mathbf{\Lambda}$ is the diagonal eigenvalue matrix with non-negative values sorted in the ascending order, i.e., $\Lambda_{ii} \leq \Lambda_{jj}$ if $i<j$, and $\mathbf{U}$ contains the associated orthonormal eigenvectors in each column.
The Graph Fourier Transform (GFT) is defined in terms of $\mathbf{U}$. 
Formally, a graph signal is a function $x: \mathcal{V} \rightarrow \mathbb{R}$ isomorphic to $\mathbb{R}^{n}$, and forms the graph signal \(\mathbf{x}\in\mathbb{R}^{n}\) consisting of node real values. Therefore, the GFT of $\mathbf{x}$ is given by $\tilde{\mathbf{x}}=\mathbf{U}^\top\mathbf{x}$, where $\tilde{x}_i$ is the spectral component of the $i$th eigenvector \cite{ortega2018graph}. 

A graph signal is smooth if connected nodes with a larger weight have more similar values \cite{dong2016learning}.
This is measured via the Laplacian's quadratic form $\mathbf{x}^\top\mathbf{L}\mathbf{x} = \sum_{i,j \in \mathcal{E}}{\mathbf{W}_{ij}(x_i-x_j)^2}$.
% \begin{equation}
% \mathbf{x}^\top\mathbf{L}\mathbf{x} = \sum_{i,j \in \mathcal{E}}{\mathbf{W}_{ij}(x_i-x_j)^2}.
% \end{equation}
Equivalently, this quadratic form can be expressed in terms of graph spectral components:
\begin{equation}
\mathbf{x}^\top\mathbf{L}\mathbf{x} = \mathbf{x}^\top\mathbf{U}\mathbf{\Lambda}\mathbf{U}^\top\mathbf{x} =
\tilde{\mathbf{x}}^\top\mathbf{\Lambda}\tilde{\mathbf{x}} =
\sum_{i=1}^n{\lambda_i \tilde{x}_i^2}\text{,}
\end{equation}
where $\lambda_i = \mathbf{\Lambda}_{ii}$ (the $i$th eigenvalue of $\mathbf{L}$) has a frequency-like interpretation \cite{dong2019learning}. 
With this notion of frequency, $h(\lambda_i)$ forms a graph filter that either amplifies or attenuates each spectral component.
The filtered graph signal
\begin{equation}
\label{gf}
\mathbf{y} = \mathbf{U}h(\mathbf{\Lambda})\mathbf{U}^\top\mathbf{x} =
h(\mathbf{L})\mathbf{x}
\end{equation}
can be characterized by applying a linear operator in terms of $\mathbf{L}$. 
Common choices for low-pass filtering includes $\mathbf{L}^{-1/2}$ or $(\mathbf{I} + \gamma \mathbf{L})^{-1}$ (for some scalar \(\gamma>0\)), and conversely $\mathbf{L}$ for high-pass filtering \cite{ortega2018graph,leus2023graph}.

\section{Kernel Multiple Graph Learning (KMGL)}

%\subsection{Problem Statement}

%\noindent \textit{Problem Statement}:

\subsubsection{Problem Statement}

We are given a (normalized) dataset $\mathcal{X} = \{\mathbf{x}_i \in \mathbb{R}^n \}_{i=1}^m$ of graph signals residing on the shared vertex set $\mathcal{V}$ and a set of node-side information $\mathcal{P} = \{\mathcal{P}_1, \cdots, \mathcal{P}_n\}$ with $\mathcal{P}_i$ denoting a prior covariate for $v_i\in\mathcal{V}$.
The objective of this study is to partition $\mathcal{X}$ into $K$ clusters $\{\mathcal{X}_k\}_{k=1}^K$ (where \(\vert \mathcal{X}_k \vert =m_k\)) and learn their associated graphs $\{\mathcal{G}_k = (\mathcal{V}, \mathcal{E}_k, \mathbf{W}_k)\}_{k=1}^K$ that best fit their partition in terms of graph signal smoothness and the node-side information. 

\subsubsection{KMGL}

To solve this problem, we select kernels \cite{hofmann2008kernel} to separate the representation of information from the algorithm.
The selection of kernels also permits the processing of different features in each cluster and consequently captures various types of relationships in each graph.
This allows us to implicitly transform each node-side information to a high-dimensional feature vector without conducting any direct computation.

Let $\mathcal{K}:\mathcal{P} \times \mathcal{P} \rightarrow \mathbb{R}$ be a symmetric positive definite kernel on the node-side information.
Based on the Aronszajn theorem \cite{aronszajn1950theory} there is a Reproducing Kernel Hilbert Space (RKHS) $\mathcal{H}$ and a feature map $\phi:\mathcal{P}\rightarrow\mathcal{H}$ such that $\mathcal{K}(\mathbf{p}_i, \mathbf{p}_j) = \langle \phi(\mathbf{p}_i), \phi(\mathbf{p}_j) \rangle_\mathcal{H}$, where $\langle \cdot , \cdot \rangle_\mathcal{H}$ denotes the inner product in the kernel space $\mathcal{H}$. 
This maps (possibly infinite) features to each node. The set of node features $\mathcal{F} = \{ \mathbf{f} \in \mathbb{R}^{n}~\vert~f_i = \phi(\mathbf{p}_i)_d,  d\in dims(\mathcal{H}) \}$
% \begin{equation}
% \mathcal{F} = \{ \mathbf{f} \in \mathbb{R}^{n}~\vert~f_i = \phi(\mathbf{p}_i)_d,  d\in dims(\mathcal{H}) \}
% \end{equation}
represents all the $n$-dimensional feature vectors associated with the kernel.
The graph signal $\mathbf{x}$ is expected to match the node-side information and may not deviate from the set of points in $\mathcal{F}$.
Thus, the first few principal components of $\mathcal{F}$ approximate $\mathbf{x}$ well.
The kernel matrix $\mathbf{K}_{ij}=\mathcal{K}(\mathbf{p}_i, \mathbf{p}_j)$ is the sample covariance of features and its eigenvectors capture these components, then the deviation is evaluated via $h(\mathbf{K})\mathbf{x}=\mathbf{K}^{-1}\mathbf{x}$ as in (\ref{gf}) for the node side \cite{pu2021kernel}. 
This transformation gets a larger effect when the graph signal has a significant projection in a direction that the feature set is less spread, \ie a small eigenvalue of $\mathbf{K}$. 
Alternatively, $h(\mathbf{K})$ can be viewed as a filter based on $\mathcal{P}$ that amplifies the signal in atypical directions that the feature set is spread.

The fitness of a graph signal $\mathbf{x}$ to the underlying graph $\mathcal{G}$ and the node-side information $\mathcal{P}$ is measured via applying a filter in terms of $\mathbf{L}$ and $\mathbf{K}$, then comparing the two signals.
We use the inner product as the similarity function as follows:
\begin{equation}
s(\mathbf{x},\hat{\mathbf{x}}) = \langle\mathbf{x},\hat{\mathbf{x}}\rangle=\mathbf{x}^\top\hat{\mathbf{x}} \text{,}
\end{equation}
where $\hat{\mathbf{x}}$ is the filtered (denoised) version of $\mathbf{x}$ such that:
\begin{gather}
    \label{eq:lincombo}
    \mathbf{x}-\hat{\mathbf{x}} = \alpha \mathbf{K}^{-1}\hat{\mathbf{x}} + \beta \mathbf{L}\hat{\mathbf{x}}\\
    \label{eq:filter}
    \Rightarrow \hat{\mathbf{x}} = \overbrace{(\mathbf{I} + \alpha \mathbf{K}^{-1} + \beta \mathbf{L})^{-1}}^{h(\mathbf{K},\mathbf{L})}\mathbf{x}
\end{gather}
% \begin{equation}
% \label{eq:lincombo}
% \mathbf{x}-\hat{\mathbf{x}} = \alpha \mathbf{K}^{-1}\hat{\mathbf{x}} + \beta \mathbf{L}\hat{\mathbf{x}}
% \end{equation}
% \begin{equation}
% \label{eq:filter}
% \Rightarrow \hat{\mathbf{x}} = \overbrace{(\mathbf{I} + \alpha \mathbf{K}^{-1} + \beta \mathbf{L})^{-1}}^{h(\mathbf{K},\mathbf{L})}\mathbf{x}
% \end{equation}
for some positive scalars $\alpha$ and $\beta$.
Eq. (\ref{eq:lincombo}) states that the difference between $\mathbf{x}$ and $\hat{\mathbf{x}}$ lies in the linear cone of $\mathbf{K}^{-1}\hat{\mathbf{x}}$ and $\mathbf{L}\hat{\mathbf{x}}$. 
The former means $\hat{\mathbf{x}}$ aligns more with the prior information, and the latter means $\hat{\mathbf{x}}$ is smoother on the underlying graph. Thus, the filter $h(\mathbf{K},\mathbf{L})$ has a low-pass behavior on the combination of the $\mathcal{G}$ and $\mathcal{P}$. 
Specifically, it becomes a typical low-pass graph filter when $\alpha=0$. 

We propose to jointly cluster the graph signals and learn multiple graphs consistently with prior node-side information. 
Our KMGL algorithm expresses the problem as finding the partition sets $\{\mathcal{X}_k\}_{k=1}^K$ and the Laplacian matrices $\{\mathbf{L}_k\}_{k=1}^{K}$ as follows:
\begin{align}
    \nonumber
    \max_{\{\mathcal{X}_k, \mathbf{L}_k\in \mathcal{L}\}_{k=1}^K, \{\hat{\mathbf{x}}_i\}_{i=1}^m} & \sum_{k=1}^{K}\sum_{\mathbf{x}\in \mathcal{X}_k}\mathbf{x}^\top\hat{\mathbf{x}} - \gamma \Vert\mathbf{L}_{k}\Vert_F^{2}\\
    \nonumber
    \textrm{s.t. }~& \mathbf{x}-\hat{\mathbf{x}} = \alpha \mathbf{K}_k^{-1}\hat{\mathbf{x}} + \beta \mathbf{L}_k\hat{\mathbf{x}};\:\forall\mathbf{x} \in \mathcal{X}_k,\:\forall k\\
    \label{eq:main}
    &\tr(\mathbf{L}_k) = n;~\forall k
\end{align}
% \begin{equation}
% \label{eq:main}
% \begin{split}
% \max_{\{\mathcal{X}_k, \mathbf{L}_k\in \mathcal{L}\}_{k=1}^K, \{\hat{\mathbf{x}}_i\}_{i=1}^m} & \sum_{k=1}^{K}\sum_{\mathbf{x}\in \mathcal{X}_k}\mathbf{x}^\top\hat{\mathbf{x}} - \gamma \Vert\mathbf{L}_{k}\Vert_F^{2}\\
% \textrm{s.t.}~& \mathbf{x}-\hat{\mathbf{x}} = \alpha \mathbf{K}_k^{-1}\hat{\mathbf{x}} + \beta \mathbf{L}_k\hat{\mathbf{x}};\:\:\forall\mathbf{x} \in \mathcal{X}_k, \forall~k\\
% &\tr(\mathbf{L}_k) = n;\:\:~\forall~k
% \end{split}
% \end{equation}
where $\mathcal{L}$ is the set of valid graph Laplacians 
\begin{equation}
\mathcal{L} = \{\mathbf{L}\in \mathbb{R}^{n\times n}~\vert~\mathbf{L}= \mathbf{L}^\top, \mathbf{L}\mathbf{1} = \mathbf{0}, \mathbf{L}_{ij}\leq 0~\forall i\ne j\}\text{.}
\end{equation}
The first term in (\ref{eq:main}) promotes the similarity of the graph signal and its filtered version.
This term helps in assigning graph signals that are more similar to the filtered version w.r.t. the underlying graph.
The hyperparameters $\alpha$ and $\beta$ control the trade-off between matching the side information and the smoothness of graph signals, respectively.
The second term in \eqref{eq:main} regularizes the graphs to have a smaller Frobinius norm.
Combined with the second constraint, they affect the sparsity of the learned graphs and avoid trivial solutions \cite{dong2016learning,pu2021kernel}.
Graphs are sparser as $\gamma \in \mathbb{R}_+$ gets larger. 
%Other regularization terms, \eg log-barrier \cite{kalofolias2016learn}, are also commonly used to impose connectedness.

\subsection{Algorithm}

The problem formulated in (\ref{eq:main}) is NP-hard.
This is because the selection of partitions affects the optimal graphs and, consequently, the objective function.
To avoid solving the problem for every possible partitioning, an iterative solution similar to K-means \cite{ikotun2022k} is proposed that increases the objective at each step.
The algorithm first partitions the dataset randomly and then iterates between two steps: i) learning a graph for each cluster, and ii) reassigning the graph signals.
This is repeated until the partitions remain the same.

\subsubsection{Fixing the cluster assignments and learning the underlying graphs and filtered signals}
Firstly, for the \(k\)th cluster, given the initial (and possibly noisy) graph signals \(\{\mathbf{x}\in\mathcal{X}_k\}\) and by fixing their assignments, we solve the following optimization problem for learning their associated filtered versions and also the \(k\)th graph Laplacian:
\begin{gather}
    \nonumber
    \{\mathbf{L}_k,\hat{\mathcal{X}}_k\} = \operatorname*{argmax}_{\mathbf{L}_k \in \mathcal{L}, \hat{\mathbf{x}}\in\hat{\mathcal{X}}_k} \sum_{\mathbf{x}\in \mathcal{X}_k}\mathbf{x}^\top\hat{\mathbf{x}} - \gamma \Vert \mathbf{L}_{k}\Vert_F^{2} \\
    \label{eq:graphsforclusters}
    \textrm{s.t.}~\forall \mathbf{x} \in \mathcal{X}_k: \mathbf{x}-\hat{\mathbf{x}} = \alpha \mathbf{K}_k^{-1}\hat{\mathbf{x}} + \beta \mathbf{L}_k\hat{\mathbf{x}},~\tr(\mathbf{L}_k) = n \text{,}
\end{gather}
% \begin{equation}
% \label{eq:graphsforclusters}
% \begin{aligned}
% &\{\mathbf{L}_k,\hat{\mathcal{X}}_k\} = \operatorname*{argmax}_{\mathbf{L}_k \in \mathcal{L}, \hat{\mathbf{x}}\in\hat{\mathcal{X}}_k} \sum_{\mathbf{x}\in \mathcal{X}_k}\mathbf{x}^\top\hat{\mathbf{x}} - \gamma \Vert \mathbf{L}_{k}\Vert_F^{2} \\
% &\textrm{s.t.}~\forall~\mathbf{x} \in \mathcal{X}_k: \mathbf{x}-\hat{\mathbf{x}} = \alpha \mathbf{K}_k^{-1}\hat{\mathbf{x}} + \beta \mathbf{L}_k\hat{\mathbf{x}},~\tr(\mathbf{L}_k) = n \text{,}\\
% \end{aligned}
% \end{equation} 
where the set \(\hat{\mathcal{X}}_k\) contains the filtered versions of graph signals for \(\mathcal{X}_k\). 
Although it is less obvious, the above problem can be formulated more similarly to the typical GL objective functions \cite{dong2016learning,pu2021kernel} as follows: 
\begin{thm}
\label{th1}
The maximization problem in (\ref{eq:graphsforclusters}) is equivalent to a joint kernel ridge regression from set $\mathcal{P}$ to $\mathcal{X}_k$ and a GL problem.
\end{thm}
\begin{proof}
We show that our problem is equivalent to:
\begin{equation}
\label{eq:dongs}
\begin{aligned}
\min_{\mathbf{L}_k \in \mathcal{L}, \mathbf{c}\in\mathcal{C}_k} & \sum_{\mathbf{x}\in \mathcal{X}_k} \Vert\mathbf{x} - \mathbf{K}_k\mathbf{c}\Vert_2^2 + \alpha \mathbf{c}^\top \mathbf{K}_k \mathbf{c} \\
& \quad \quad +\beta \mathbf{c}^\top \mathbf{K}_k\mathbf{L}_k\mathbf{K}_k\mathbf{c}+\gamma\Vert\mathbf{L}_{k}\Vert_F^{2} \\
\textrm{s.t. }~& \tr(\mathbf{L}_k) = n,
\end{aligned}
\end{equation} 
where for any \(\mathbf{c}\in\mathcal{C}_k\) there exist only one filtered signal \(\hat{\mathbf{x}}\in\hat{\mathcal{X}}_k\) such that $\hat{\mathbf{x}} = \mathbf{K}_k\mathbf{c}$ as proposed in \cite{pu2021kernel}.
We start by writing (\ref{eq:dongs}) in terms of the filtered signals $\{\hat{\mathbf{x}}\in\hat{\mathcal{X}}_k\}$. 
Note that since $\mathbf{K}_k$ is positive definite, $\mathbf{c}^\top\mathbf{K}_k\mathbf{c} = \hat{\mathbf{x}}^\top\mathbf{K}^{-1}_k\hat{\mathbf{x}}$ and we have:
\begin{align}
\nonumber
\min_{\mathbf{L}_k \in \mathcal{L}, \hat{\mathcal{X}}_k} & \sum_{\mathbf{x}\in \mathcal{X}_k} \Vert\mathbf{x} - \hat{\mathbf{x}}\Vert_2^2 + \alpha \hat{\mathbf{x}}^\top \mathbf{K}^{-1}_k \hat{\mathbf{x}} + \beta \hat{\mathbf{x}}^\top\mathbf{L}_k\hat{\mathbf{x}} + \gamma\Vert\mathbf{L}_{k}\Vert_F^{2} \\
\label{eq:donggraphlearning}
\textrm{s.t. }~& \tr(\mathbf{L}_k) = n.
\end{align}
% \begin{equation}
% \begin{aligned}
% \label{eq:donggraphlearning}
% \min_{\mathbf{L}_k \in \mathcal{L}, \hat{\mathcal{X}}_k} & \sum_{\mathbf{x}\in \mathcal{X}_k} \Vert\mathbf{x} - \hat{\mathbf{x}}\Vert_2^2 + \alpha \hat{\mathbf{x}}^\top \mathbf{K}^{-1}_k \hat{\mathbf{x}} + \beta \hat{\mathbf{x}}^\top\mathbf{L}_k\hat{\mathbf{x}} + \gamma\Vert\mathbf{L}_{k}\Vert_F^{2} \\
% \textrm{s.t. }~& \tr(\mathbf{L}_k) = n.\\
% \end{aligned}
% \end{equation} 
The above problem is separable in each $\hat{\mathbf{x}}$ and we start by minimizing over them as follows:
\begin{equation}
\label{x_hat}
\hat{\mathbf{x}} = \operatorname*{argmin}_{\hat{\mathbf{x}}} \underbrace{\Vert\mathbf{x} - \hat{\mathbf{x}}\Vert_2^2 + \alpha \hat{\mathbf{x}}^\top \mathbf{K}^{-1} _k\hat{\mathbf{x}} + \beta \hat{\mathbf{x}}^\top\mathbf{L}_k\hat{\mathbf{x}}}_{f(\hat{\mathbf{x}})}
\end{equation}
that is convex and differentiable with a gradient:
\begin{equation}
\label{grad}
\nabla_{\hat{\mathbf{x}}}f(\hat{\mathbf{x}}) = -(\mathbf{x} - \hat{\mathbf{x}}) +\alpha \mathbf{K}_k^{-1}\hat{\mathbf{x}} + \beta \mathbf{L}_k\hat{\mathbf{x}}.
\end{equation}
Putting the gradient in (\ref{grad}) to zero results in (\ref{eq:lincombo}). 
Then, by substituting (\ref{eq:lincombo}) into (\ref{x_hat}), one can write:
\begin{equation}
\notag
\begin{split}
f(\hat{\mathbf{x}})&= \Vert\mathbf{x} - \hat{\mathbf{x}}\Vert_2^2 + \hat{\mathbf{x}}^\top\left(\alpha  \mathbf{K}^{-1} _k\hat{\mathbf{x}} + \beta\mathbf{L}_k\hat{\mathbf{x}}\right) \\ 
&= ||\mathbf{x} - \hat{\mathbf{x}}||_2^2 + \hat{\mathbf{x}}^\top(\mathbf{x} - \hat{\mathbf{x}})\\
&= \mathbf{x}^\top(\mathbf{x}-\hat{\mathbf{x}}) = ||\mathbf{x}||_2^2 - \mathbf{x}^\top\hat{\mathbf{x}}.\\
\end{split}
\end{equation}
Next, objective (\ref{eq:donggraphlearning}) turns to:
\begin{equation}
\begin{aligned}
\min_{\mathbf{L}_k \in \mathcal{L}, \hat{\mathcal{X}}_k} & \sum_{\mathbf{x}\in \mathcal{X}_k}\Vert\mathbf{x}\Vert_2^2 -\mathbf{x}^\top\hat{\mathbf{x}} + \gamma \Vert\mathbf{L}_{k}\Vert_F^{2} \\
\textrm{s.t. }~&\forall \mathbf{x} \in \mathcal{X}_k: \mathbf{x}-\hat{\mathbf{x}} = \alpha \mathbf{K}_k^{-1}\hat{\mathbf{x}} + \beta \mathbf{L}_k\hat{\mathbf{x}} \\
& \tr(\mathbf{L}_k) = n.\\
\end{aligned}
\end{equation} 
The term $\Vert\mathbf{x}\Vert_2^2$ has no bearing on the minimization, and a sign change in objective results in the maximization (\ref{eq:graphsforclusters}).
\end{proof}

Based on Theorem \ref{th1} and the approach in \cite{pu2021kernel}, we solve (\ref{eq:graphsforclusters}) by applying a BCD scheme on objective (\ref{eq:donggraphlearning}) that iteratively filters the graph signals by Eq. (\ref{eq:filter})
and then solves for:
\begin{equation}
\begin{aligned}
\label{eq:graphlearning}
\mathbf{L}_k=\operatorname*{argmin}_{\mathbf{L}_k \in \mathcal{L}} & \sum_{\forall\hat{\mathbf{x}}\in \hat{\mathcal{X}}_k} {\beta \hat{\mathbf{x}}^\top\mathbf{L}_k\hat{\mathbf{x}}} + \gamma||\mathbf{L}_{k}||_F^{2} \\
\textrm{s.t. }~& \tr(\mathbf{L}_k) = n,
\end{aligned}
\end{equation} 
which is a GL problem in the typical Laplacian quadratic form \cite{dong2016learning,kalofolias2016learn} and can be solved by convex optimization techniques \cite{diamond2016cvxpy}.

\algrenewcommand\algorithmicrequire{\textbf{Input:}}
\algrenewcommand\algorithmicensure{\textbf{Output:}}

\begin{algorithm}[t]
\caption{: KMGL}\label{alg:cap}
\begin{algorithmic}[1]
\Require Graph signals $\mathcal{X}$, number of clusters $K$, Kernel matrices $\{\mathbf{K}_k\}_{k=1}^{K}$, hyperparameters $\alpha, \beta, \gamma$, tolerance $\epsilon$
\Ensure Partition set $\{\mathcal{X}_k\}_{k=1}^K$, graph Laplacians $\{\mathbf{L}_k\}_{k=1}^K$, filtered signals $\{ \forall~\mathbf{x} \in \mathcal{X}:\hat{\mathbf{x}} \}$
\State \textbf{Initialization}: Randomly partition $\mathcal{X}$ into $K$ clusters.
\Repeat
\For{$k=1:K$} \Comment{GL and filtering (denoising)}
\Repeat
\State Filter every $\mathbf{x} \in \mathcal{X}_k$ with $h(\mathbf{K}_k, \mathbf{L}_k)$ in (\ref{eq:filter})
\State Learn the graph Laplacian $\mathbf{L}_k$ via Eq. (\ref{eq:graphlearning})
\Until  $\mathbf{L}_k$ converges w.r.t. a tolerance $\epsilon$
\EndFor
\For{$\mathbf{x} \in \mathcal{X}$} \Comment{Refining the clusters}
\State Update the cluster of $\mathbf{x}$ via Eq. (\ref{eq:clusterupdate})
\EndFor
\Until clusters are unchanged
\end{algorithmic}
\end{algorithm}

\subsubsection{Assigning the graph signals to their associated clusters by fixing the underlying graphs and filtered signals.}
In the second step of the algorithm, we refine the partitions by fixing the graphs and assigning each graph signal to the most compatible cluster. 
Let $\hat{\mathcal{I}}(\mathbf{x}) = \{ \hat{\mathbf{x}}_k~\vert~\hat{\mathbf{x}}_k = h(\mathbf{K}_k,\mathbf{L}_k)\mathbf{x},k=1,\dots,K \}$ be the set of filtered graph signals of $\mathbf{x}$ over all graphs. We specify the assignment of \(\mathbf{x}\), \ie \(i(\mathbf{x})\), as follows:
\begin{equation}
\label{eq:clusterupdate}
i(\mathbf{x}) = \operatorname*{argmax}_{k:~\hat{\mathbf{x}}_k \in \hat{\mathcal{I}}(\mathbf{x})}{ \mathbf{x}^\top\hat{\mathbf{x}}_k}\text{.}
\end{equation}
Then, the partitions are refined such that $\mathcal{X}_k = \{ \mathbf{x} \in \mathcal{X}~\vert~i(\mathbf{x}) = k \}$ for $k=1,\dots,K$.

These two steps continue alternatively until getting convergence, \eg clusters are unchanged. The proposed KMGL algorithm is summarized in Algorithm \ref{alg:cap}.

% \begin{figure*}[t]
%     \centering
%     \begin{minipage}{\columnwidth}
%     \begin{subfigure}[b]{\textwidth}
%         \centering
%          \includegraphics[width=\textwidth]{CAR-K.eps}
%          \caption{SNR is fixed at 15 dB}
%      \end{subfigure}
%      \begin{subfigure}[b]{\textwidth}
%          \centering
%          \includegraphics[width=\textwidth]{CAR-SNR.eps}
%          \caption{Results for 3 clusters}
%      \end{subfigure}
%      \caption{Clustering performance of KMGL compared to K-graphs, GLMM, and K-means based on CAR when (a) $K$ and (b) SNR increases. $\alpha=\beta=10^{-2}$, $\gamma=10^{-4}$.}
%     \label{fig:car}
%     \end{minipage}\hfill
%         \begin{minipage}{\columnwidth}
%     \begin{subfigure}[b]{\textwidth}
%         \centering
%          \includegraphics[width=\textwidth]{APS-K.eps}
%          \caption{SNR is fixed at 15 dB}
%      \end{subfigure}
%      \begin{subfigure}[b]{\textwidth}
%          \centering
%          \includegraphics[width=\textwidth]{APS-SNR.eps}
%          \caption{Results for 3 clusters}
%      \end{subfigure}
%      \caption{GL performance of KMGL compared to K-graphs, and GLMM based on APS when (a) $K$ and (b) the SNR increases. $\alpha=\beta=10^{-2}$, $\gamma=10^{-4}$.}
%     \label{fig:snr}
%     \end{minipage}
% \end{figure*}

The next theorem states some properties about the convergence of the proposed KMGL method.

\begin{thm}
The KMGL algorithm converges in a finite number of iterations.
\end{thm}
\begin{proof}
For each algorithm step, the objective function in (\ref{eq:main}) is non-decreasing.
This is because in the first step when the graphs are updated via Eq. (\ref{eq:graphsforclusters}), the maximization is over the same terms, and thus the new graphs will not decrease the objective. 
Precisely, the optimization problem \eqref{eq:graphsforclusters} is biconvex \cite{pu2021kernel}, and therefore, utilizing BCD reaches unique solutions for each subproblem which guarantees to reach a stationary point \cite{beck2013convergence}.
In the second step, when the clusters are refined, we directly increase the term $\mathbf{x}^\top\hat{\mathbf{x}}$ for each graph signal, or it remains the same.
Since $\Vert\mathbf{L}_k\Vert_F^2$ is fixed, the objective is also non-decreasing here.
Lastly, there are finite assignments of $m$ graph signals to $K$ clusters, and consequently, the algorithm has to converge \cite{ikotun2022k}.
\end{proof}

\section{Experiments}

In this section, the performance of the KMGL algorithm on numerical data is evaluated and compared to the GLMM \cite{maretic2020graph}, and K-graphs \cite{araghi2019k} methods.
The K-means algorithm is also added as a baseline to represent a model without knowledge of the underlying graphs.
We draw random Erdos–Renyi graphs of $n$ nodes with a (binary) connection probability of $p((v_i,v_j) \in \mathcal{E}) =0.3$ for $i\ne j$.
Edge weights are normalized such that the sum of weights is $n$. 
Similar to \cite{pu2021kernel}, graph signals of the \(k\)th cluster are generated according to $\forall \mathbf{x}\in\mathcal{X}_k: \mathbf{x} \sim \mathcal{N}(\mathbf{0}, \mathbf{K}_k + \sigma_\epsilon \mathbf{I})$, where $\mathbf{K}_k = (\mathbf{I} + \eta \mathbf{L}_k)^{-1}$ is the kernel matrix corresponding to the \(k\)th cluster and $\sigma_\epsilon$ is the noise-level.
This common choice of kernel leads to globally smooth signals \cite{pu2021kernel}. 
We select $\eta = 10$ for the experiments by performing a grid search on the training data.
The models are examined on different $\sigma_\epsilon$ and number of clusters $K$. 
We select the noise level such that the Signal-to-Noise Ratio
\begin{equation}
\text{SNR} = 
% 10 \log_{10}(\frac{1}{n\sigma_\epsilon^2}\mathbb{E}[||\mathbf{x}||_2^2]) = 
10 \log_{10}\left(\frac{1}{Kn\sigma_\epsilon^2}\sum_{c=1}^K \tr(\mathbf{K}_c)\right)
\end{equation}
is varied uniformly.

Models are evaluated based on their clustering performance and the quality of learned graphs. 
The former is measured via Clustering Accuracy Ratio (CAR), which finds the best map between the cluster indices of samples and the partitions, then measures the number of correctly clustered signals to their total number \cite{araghi2019k}.
The quality of learned graphs is evaluated by Average Precision Score (APS), where the ability of the model to detect the presence of edges is considered \cite{pu2021kernel}.
%APS automatically varies the threshold that weights are declared present and then compares them with the ground truth \cite{pu2021kernel}.

\begin{figure}
    \centering
    \includegraphics[width=\columnwidth]{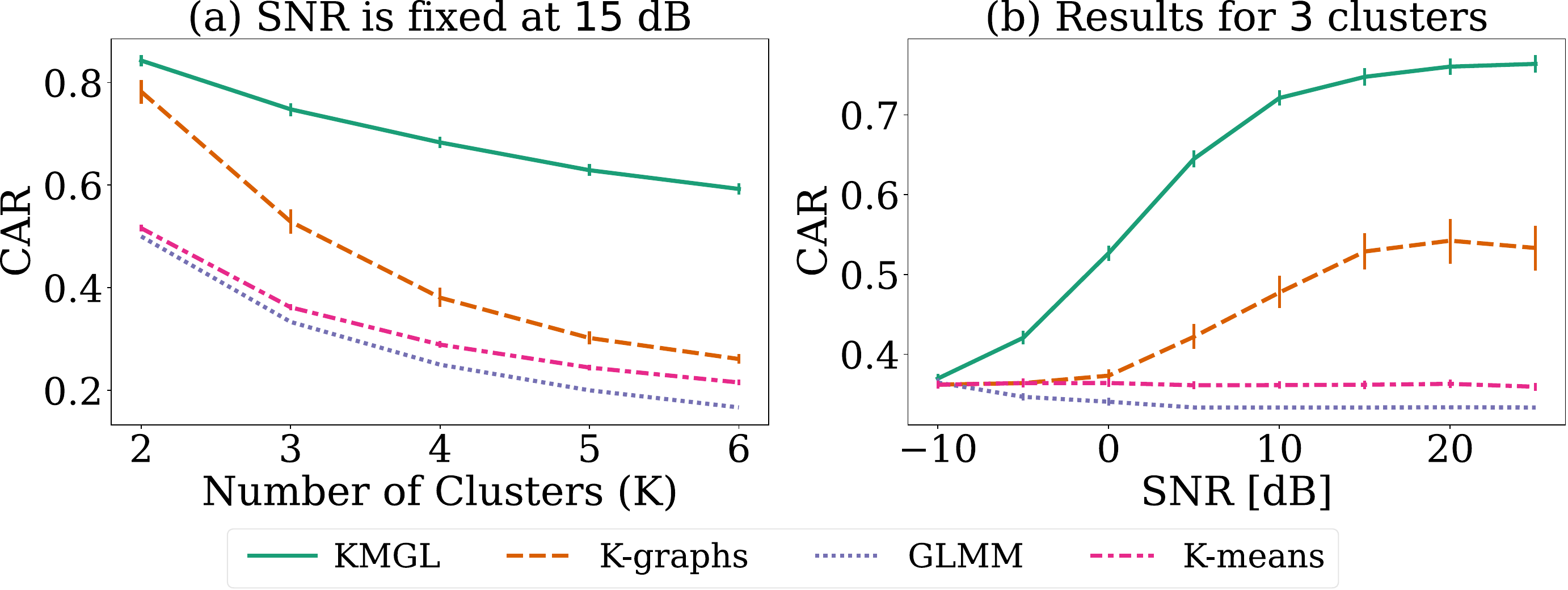}
    \caption{Clustering performance of KMGL compared to K-graphs, GLMM, and K-means based on CAR when (a) $K$ and (b) SNR increases. $\alpha=\beta=10^{-2}$, $\gamma=10^{-4}$.}
    \label{fig:car}
\end{figure}

\begin{figure}
    \centering
    \includegraphics[width=\columnwidth]{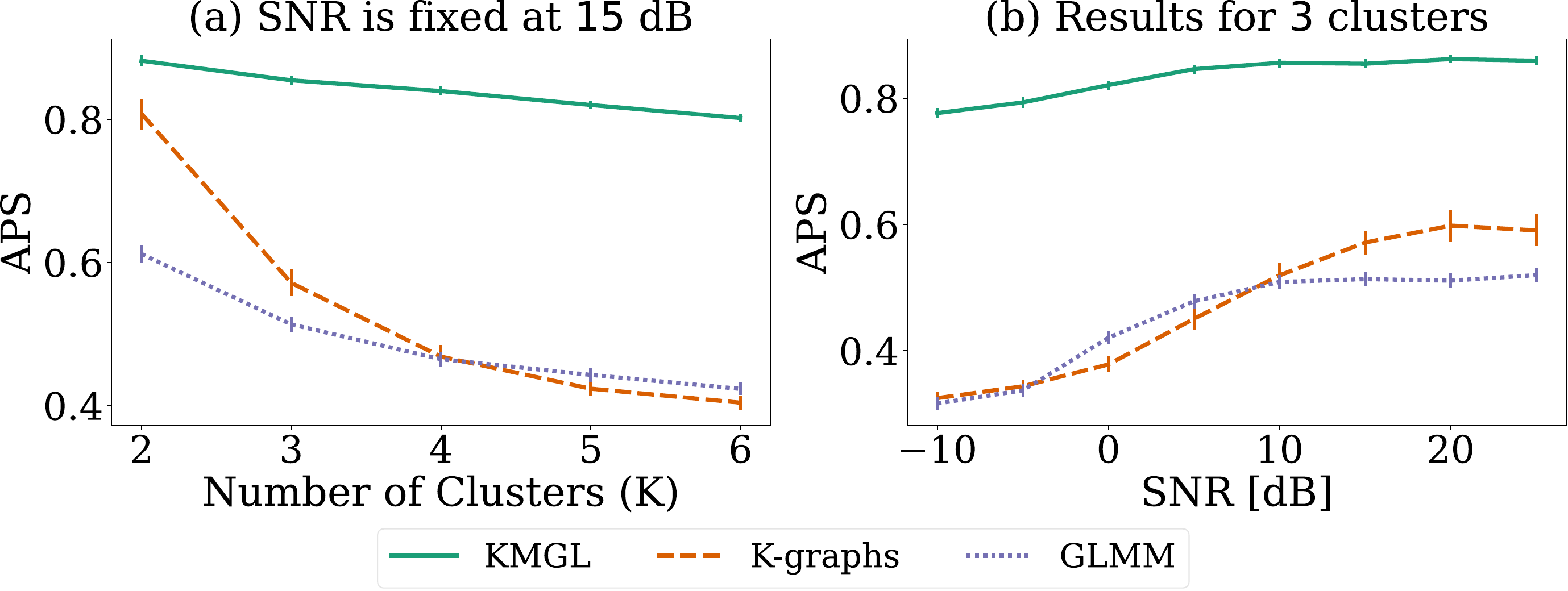}
    \caption{GL performance of KMGL compared to K-graphs, and GLMM based on APS when (a) $K$ and (b) the SNR increases. $\alpha=\beta=10^{-2}$, $\gamma=10^{-4}$.}
    \label{fig:snr}
\end{figure}

The datasets have $m=500$ graph signals, sampled equally from $K\in\{2,3,4,5\}$ different graphs with $n=20$ nodes. 
We perform preliminary tests for each model to tune the hyperparameters and then keep them intact through the experiments.
Furthermore, we restrict our model so that $\alpha = \beta = 10^{-2}$ equally prioritizes smoothness and side information. 
In the results, each data point consists of $50$ independent realizations.
The compared models are applied $10$ times for each realization and are evaluated on their best try, determined via their objective function, while the proposed algorithm is applied only once. 

Fig. \ref{fig:car} displays the effective clustering performance of the KMGL algorithm. 
Fig. \ref{fig:car}(a) shows that KMGL is more robust to a high number of clusters, and Fig. \ref{fig:car}(b) shows KMGL's robustness against noise compared to the previous methods.
Fig. \ref{fig:snr} shows the ability of the compared models to recover and learn the graphs, even in a high number of clusters in Fig. \ref{fig:snr}(a) and a high amount of noise in Fig. \ref{fig:snr}(b).
Similar to Fig. \ref{fig:car}, KMGL outperforms the compared models both in Fig. \ref{fig:snr}(a) when the number of clusters is increased and Fig. \ref{fig:snr}(b) when the noise-rejection behavior is studied.
It is worth mentioning that the ability of the model to recover the graphs even in high noise levels is an effective advantage of exploiting kernel metrics that was also seen in \cite{pu2021kernel}. 
This further shows the benefits of incorporating side information.
%As shown in Fig. \ref{fig:car} and \ref{fig:snr}, K-graphs perform more robustly than GLMM, probably due to utilizing a hard clustering strategy.

\section{conclusion}

In this letter, we introduced the KMGL algorithm, which incorporates node-side information in clustering graph signals and learning multiple graphs.
We used kernels to represent this node-side information and built a framework that uses filters to model the relationship between the data and the underlying graphs.
We solved the optimization problem associated with KMGL using the BCD method, and we proved its convergence.
Our experiments have shown that KMGL outperforms existing methods, especially when dealing with high levels of noise and a large number of clusters. 
The theoretical guarantees and experiments underscore the potential value of KMGL in real-world applications.

%The experimental results showed the effectiveness of this method, both in terms of clustering and GL.

% \section{Acknowledgment}
% This work was (partially) supported by the center Hi! PARIS.

\clearpage

\bibliographystyle{ieeetr}
\bibliography{references}

\clearpage
\appendix
\subsection{Clustering and Learning with Missing Data}

In this section, we extend the KMGL algorithm to support partially observed graph signals. 
Specifically, we change how each graph signal is filtered by first recovering its missing values. 
To this end, we adopt an iterative approach for graph signal reconstruction. 
The asymptotic behavior of the resulting algorithm is studied and we conduct numerical experiments to measure its effectiveness. 

Let $\mathcal{B}$ denote the observed subspace of the graph signal $\mathbf{x}$. Then, the downsampling operator $\mathbf{J}: \mathbb{R}^n \rightarrow \mathcal{B}$ maps the partially observed signal to this space. 
Moreover, $\mathbf{M} = \mathbf{J}^\top\mathbf{J}$ represents the downsampling and then upsampling operation where $\mathbf{M}$ is a diagonal masking matrix with $M_{ii} = 1$ if we observe the $i$th component of $\mathbf{x}$ and zero otherwise. 
$\mathbf{M}\mathbf{x}$ interpolates the missing values of $\mathbf{x}$ with zeroes, however, this incorporates neither the graph structure nor the side information.

To extend the KMGL algorithm, we need to revisit how each graph signal $\mathbf{x}$ relates to its filtering $\hat{\mathbf{x}}$. 
Previously, the low-pass filter $\mathbf{S} = h(\mathbf{K}, \mathbf{L})$ related the two in (\ref{eq:filter}).
Since $\mathbf{x}$ is partially observed,
\begin{align}
    \label{eq:reconx}
    &     \mathbf{x}^{1} = \mathbf{M}\mathbf{x} \notag \\
    &     \hat{\mathbf{x}}^{t} = \mathbf{S}\mathbf{x}^{t}  \notag\\
    &      \mathbf{x}^{t+1} = \hat{\mathbf{x}}^t + \mathbf{M}(\mathbf{x}^1-\hat{\mathbf{x}}^t)
\end{align}
iteratively recovers its missing values as suggested in \cite{narang2013localized}.
The first line initially interpolates $\mathbf{x}$ with zero.
The second line applies a low-pass filter to update the missing nodes based on the others. 
The structure of the graph and the side information dictate how missing values relate to others.
The last line ensures that $\mathbf{x}^t$ remains unchanged in the observed nodes. 
Lastly, the similarity of the reconstructed graph signal $\mathbf{x}^t$ with its low-passed filtering $\hat{\mathbf{x}}^t$ is compared in $\mathcal{B}$:
\begin{equation}
    s(\mathbf{x}, \hat{\mathbf{x}}^t)=(\mathbf{J}\mathbf{x}^t)^\top(\mathbf{J}\hat{\mathbf{x}}^t) = 
    % \mathbf{x}_t^T\mathbf{M}\hat{\mathbf{x}}_t =
    \mathbf{x}^\top\mathbf{M}\hat{\mathbf{x}}^t.
\end{equation}

\begin{thm}
As the number of iterations $t$ increases, $\hat{\mathbf{x}}^t$ has the asymptotic solution of $(\mathbf{M} + \alpha\mathbf{K}^{-1} + \beta \mathbf{L})^{-1}\mathbf{M}\mathbf{x} = \hat{\mathbf{x}}$.
\end{thm}

\begin{proof}
The iterations in (\ref{eq:reconx}) converge to a fixed point if $\mathbf{S}$ is a non-expansive operator \cite{narang2013localized}. 
Since $\alpha\mathbf{K}^{-1} + \beta \mathbf{L}$ is positive definite, all the eigenvalues of $\mathbf{S}$ have the absolute value of less than one. 
Consequently, $\mathbf{S}$ is non-expansive and $\hat{\mathbf{x}}_t$ converges.
In the converged point $\hat{\mathbf{x}} = \hat{\mathbf{x}}^t = \hat{\mathbf{x}}^{t+1}$ we have
\begin{align}
\label{eq:reconxasymp}
    & \hat{\mathbf{x}} = \mathbf{S}(\hat{\mathbf{x}} + \mathbf{M}(\mathbf{x}_1 - \hat{\mathbf{x}})) = \mathbf{S}(\hat{\mathbf{x}} + \mathbf{M}\mathbf{x} - \mathbf{M}\hat{\mathbf{x}}) \Rightarrow \notag \\
    & \mathbf{M}(\mathbf{x}-\hat{\mathbf{x}}) = \mathbf{S}^{-1}\hat{\mathbf{x}} - \hat{\mathbf{x}} = \alpha\mathbf{K}^{-1}\hat{\mathbf{x}} + \beta \mathbf{L}\hat{\mathbf{x}} \Rightarrow \notag \\
    & (\mathbf{M} + \alpha\mathbf{K}^{-1} + \beta \mathbf{L})^{-1}\mathbf{M}\mathbf{x} = \hat{\mathbf{x}}
\end{align}
which completes the proof.
\end{proof}

We remark that (\ref{eq:reconxasymp}) also results from solving the following convex problem as shown in \cite{pu2021kernel}:
\begin{equation}
    \min_{\hat{\mathbf{x}}}{\Vert\mathbf{M}(\mathbf{x}-\hat{\mathbf{x}})\Vert_2^2 + \alpha \hat{\mathbf{x}}^\top \mathbf{K}^{-1}\hat{\mathbf{x}} + \beta \hat{\mathbf{x}}^\top\mathbf{L}\hat{\mathbf{x}}}.
\end{equation}

To summarize, the Alg. \ref{alg:cap} is generalized to support partially observed graph signals as follows.
In line 5, every signal is now filtered via (\ref{eq:reconxasymp}), where $\mathbf{M}$ is the diagonal masking matrix associated with $\mathbf{x}$.
The Laplacian and the kernel also relate to each signal's cluster.
In line 10, the signals are reassigned based on:
\begin{equation}
    \label{eq:maskedclusterupdate}
    i(\mathbf{x}) = \operatorname*{argmax}_{k:\hat{\mathbf{x}}_k \in \hat{\mathcal{I}}(\mathbf{x})}{\quad \mathbf{x}^\top\mathbf{M}\hat{\mathbf{x}}_k}
\end{equation}
where
\begin{equation}
    \notag
    \hat{\mathcal{I}} = \{ \hat{\mathbf{x}}_k~\vert~(\mathbf{M} + \alpha\mathbf{K}_k^{-1} + \beta \mathbf{L}_k)^{-1}\mathbf{M}\mathbf{x} = \hat{\mathbf{x}}_k, 1\leq k \leq |\mathcal{C}|\}
\end{equation}
is similar to the previous set but with the filters in (\ref{eq:reconxasymp}).

The rest of this section discusses the validity of the proposed algorithm in numerical experiments. 
Each entry of each graph signal has a $(1-r)$ probability of missing where $r$ is the missing rate.
Hence, the element of the diagonal mask matrix is generated via $M_{ii} \sim \Bernoulli(1-r)$.
For the compared models, the missing values are set to their statistical expectation, \ie zero, as this is a natural and unbiased setting in practice. 

\begin{figure}
    \centering
    \includegraphics[width=\columnwidth]{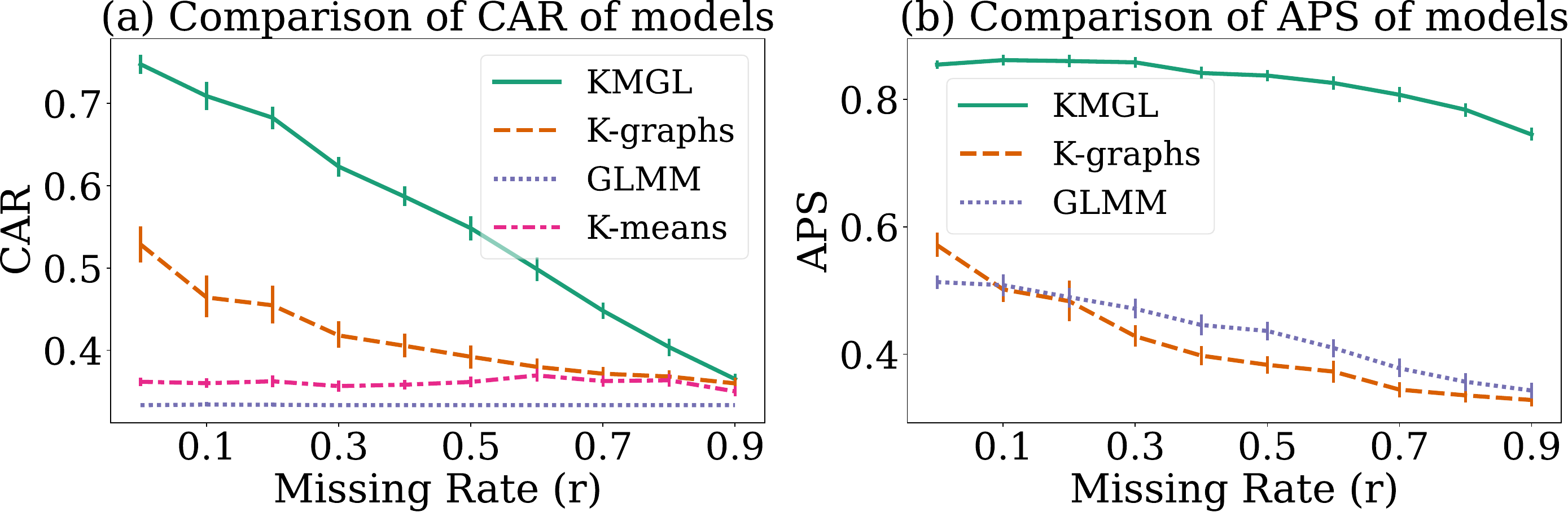}
    \caption{Performance of KMGL algorithm on partially observed signals compared to K-graphs, GLMM, and K-means in terms of (a) clustering and (b) learning graphs. SNR=15 and 3 clusters.}
    \label{fig:missingdata}
\end{figure}

% \begin{figure}
%     \centering
%     \begin{subfigure}[b]{\columnwidth}
%         \centering
%          \includegraphics[width=\textwidth]{CAR-r.eps}
%          \subcaption{Comparison of clustering accuracy ratio (CAR) of models}
%      \end{subfigure}
%     \begin{subfigure}[b]{\columnwidth}
%         \centering
%          \includegraphics[width=\textwidth]{APS-r.eps}
%          \subcaption{Comparison of average precision score (APS) of models}
%      \end{subfigure}
%      \caption{Performance of KMGL algorithm on partially observed signals compared to K-graphs, GLMM, and K-means in terms of (a) clustering and (b) learning graphs. SNR=15 and 3 clusters.}
% \label{fig:missingdata}
% \end{figure}

The performance of models is evaluated based on their Clustering Accuracy Ratio (CAR) and Average Precision Score (APS) of the recovered graphs.
The metrics are plotted against the missing rate as it changes uniformly between zero and one.
Fig. \ref{fig:missingdata} summarizes the results and shows the higher robustness of KMGL.
Each data point in the figures is the average of $20$ independent realizations with different graphs and data.

\end{document}